\documentclass[12pt]{article}
\usepackage{epsfig}
\usepackage{color}
\usepackage{amssymb,amsmath}

\newcommand{\bea}{\begin{eqnarray}}
\newcommand{\eea}{\end{eqnarray}}

 \makeatletter
\def\alt{\mathrel{\mathpalette\gl@align<}}
\def\agt{\mathrel{\mathpalette\gl@align>}}
\def\gl@align#1#2{\lower.6ex\vbox{\baselineskip\z@skip\lineskip\z@
\ialign{$\m@th#1\hfil##\hfil$\crcr#2\crcr\sim\crcr}}} \makeatother

\begin{document}
\begin{flushright}
KEK-TH-1303
\end{flushright}
\vspace*{1.0cm}

\begin{center}
\baselineskip 20pt 
{\Large\bf 
Classically conformal $B-L$ extended Standard Model 
}
\vspace{1cm}

{\large 
Satoshi Iso\footnote{satoshi.iso@kek.jp}, 
Nobuchika Okada\footnote{okadan@post.kek.jp} 
and 
Yuta Orikasa\footnote{orikasa@post.kek.jp}
} \vspace{.5cm}

{\baselineskip 20pt \it
Institute of Particle and Nuclear Studies, \\ 
High Energy Accelerator Research Organization (KEK)  \\
and \\
Department of Particles and Nuclear Physics, \\
The Graduate University for Advanced Studies (SOKENDAI), 
\\
1-1 Oho, Tsukuba, Ibaraki 305-0801, Japan} 

\vspace{.5cm}

\vspace{1.5cm} {\bf Abstract}
\end{center}

Under a hypothesis of classically conformal theories, 
we investigate the minimal $B-L$ extended Standard Model, 
which naturally provides the seesaw mechanism for 
explaining tiny neutrino masses. 
In this setup, the radiative gauge symmetry breaking is 
successfully realized in a very simple way: 
The $B-L$  gauge symmetry is broken through the conformal anomaly 
induced by quantum corrections in the Coleman-Weinberg potential. 
Associated with this $B-L$  symmetry breaking, the Higgs mass parameter 
is dynamically generated, by which the electroweak symmetry 
breaking is triggered. 
We find that a wide range of parameter space can satisfy 
both the theoretical and experimental requirements.

\thispagestyle{empty}

\newpage

\addtocounter{page}{-1}
\setcounter{footnote}{0}
\baselineskip 18pt
{\bf Introduction}.---
The stability of the electroweak scale, in other words, 
the gauge hierarchy problem is one of the most important issues 
in the Standard Model (SM), which has been motivating us to 
seek new physics beyond the SM for decades. 
The problem originates from the quadratic divergence 
in quantum corrections to the Higgs self energy, which 
should be canceled by the Higgs mass parameter  
with extremely high precision when the cutoff scale is much 
higher than the electroweak scale, say, the Planck scale. 
The most popular new physics scenario which offers 
the solution to the gauge hierarchy problem is 
the supersymmetric extension of the SM 
where no quadratic divergence arises by virtue of supersymmetry.

Because of the chiral nature of the SM, the SM Lagrangian 
at the classical level possesses the conformal invariance 
except for the Higgs mass term closely related to 
the gauge hierarchy problem. 
Bardeen has argued  
\cite{Bardeen} that once the classical conformal invariance 
 and its minimal violation  by quantum anomalies are imposed on the SM, 
it can be free from 
the quadratic divergences and hence the gauge hierarchy problem.
If the mechanism  really works, we can directly
interpolate the electroweak scale and the Planck scale.

As was first demonstrated by Coleman and Weinberg \cite{CW} 
for the U(1) gauge theory with a massless scalar, 
the classical conformal invariance is broken by quantum corrections 
in the Coleman-Weinberg effective potential and the mass scale 
is generated through the dimensional transmutation. 
It is a very appealing feature of this scheme that associated 
with this conformal symmetry breaking, the gauge symmetry is also 
broken and the Higgs boson arises as a pseudo-Nambu-Goldstone boson 
whose mass has a relationship with the gauge boson mass  
and hence predictable (when only the gauge coupling is considered). 

It was tempting to apply this Coleman-Weinberg mechanism 
to the SM Higgs sector, and the upper bound on the Higgs boson 
mass was found as $m_h \lesssim 10$ GeV under the assumption 
for a top quark mass $m_t \lesssim m_Z$ \cite{CW}. 
Unfortunately, this possibility is already excluded 
by the current experimental bound on the Higgs boson mass 
$m_h > 114.4$ GeV \cite{LEP2} and the top quark mass measurement 
$m_t = 172.4$ GeV \cite{Tevatron}. 
In addition, there is an obvious theoretical problem 
that the Coleman-Weinberg effective potential in the SM 
becomes unbounded from below for $m_t > m_Z$ \cite{fujikawa}. 
Therefore, in order to pursue this scheme, 
it is necessary to extend the SM. 
In fact, it has been found \cite{Hempfling, Espinosa, Chang, Foot, MaNi}
that when the SM is extended to include an additional scalar, 
phenomenologically viable models can be obtained. 

The reason that  the Higgs boson mass was predicted to
be lighter than $10$ GeV
in the original Coleman-Weinberg model is as follows. 
The effective potential in a classically conformal theory is given by
\begin{equation}
V_{eff}=\frac{\lambda h^4}{4} + B h^4 \left(
\ln \left(\frac{h^2}{\langle h \rangle^2}\right)-\frac{25}{6} \right)
\end{equation}
 where
the renormalization scale is taken at $\langle h \rangle$
and the coefficient $B$ for the SM is given by
\begin{equation}
B=\frac{3}{64\pi^2}
\left(\frac{3g^4+2g^2g^{\prime 2} +g^{\prime 4}}{16}-g_t^4 \right)
\end{equation}
where $g$, $g^\prime$ and  $g_t$  
are the gauge couplings and the top Yukawa coupling, respectively.
Here we have neglected the contributions of the other fields
including the scalar itself. 
The coupling constant is renormalized as 
$V^{(4)}|_{h=\langle h \rangle}=6 \lambda$. 
Then the effective potential must satisfy
$V'|_{h=\langle h \rangle}=0$ 
and the renormalized coupling constant $\lambda$ 
is related to $B$ as 
\begin{equation}
\lambda = \frac{44}{3} B.
\label{0-1relation}
\end{equation}
The Higgs boson mass is given 
by 
\begin{equation}
m_{h}^2=V_{eff}^{(2)}|_{h=\langle h \rangle}
  = 8B \langle H \rangle^2
=\frac{6}{11}
\lambda \langle h \rangle^2.
\label{CWmass}
\end{equation}  
In order for the perturbative calculation to be valid,
the coupling constant $\lambda$ must be balanced with the
1-loop coefficient $B$. If the top Yukawa coupling
were neglected, the 
dominant contribution to $B$ would come from the gauge couplings
and hence $\lambda$ could not be so large.
This gave the upper bound for the Higgs boson mass in the 
original Coleman-Weinberg scenario.
At present we know that because of 
the large top quark contribution $B$ is negative
and  the Coleman-Weinberg potential does not have a stable
minimum with a positive $m_{h}^2$.
The relation (\ref{CWmass}) should be compared with 
a similar relation  for the classical potential with
a negative mass squared term ($\mu^2 <0$)
\begin{equation} 
V=\frac{\lambda}{4} h^4 + \frac{\mu^2}{2} h^2.
\label{classicalpotential}
\end{equation}
In this case, the Higgs boson mass is given by
\begin{equation}
m_{h}^2 
=  2|\mu^2|
=2 \lambda \langle h \rangle^2,
\label{SMmass}
\end{equation}
and the coefficient is about 4 times larger
than (\ref{CWmass}). Furthermore the coupling constant $\lambda$
can become relatively larger than before so long as
it does not diverge up to the ultra-violet cutoff scale,
which we take at the Planck scale.

If both of the classical mass term  $\mu^2$ and the radiative
corrections $B$ are present,
the relation (\ref{0-1relation}) is modified to
\begin{equation}
\lambda = \frac{44}{3} B - \frac{\mu^2}{\langle h \rangle^2}
\end{equation}
and
the Higgs boson mass is given by
\begin{equation}
m_{h}^2 
= \frac{6 }{11} \lambda \langle h \rangle^2 +  \frac{16}{11} |\mu^2|
=\left( 2 \lambda - \frac{64}{3} B\right) \langle h \rangle^2.
\label{CWSMmass}
\end{equation}
This generalizes  the two extremal cases
(\ref{CWmass}) and  (\ref{SMmass}). 


The current experimental observations, in particular, 
the solar and atmospheric neutrino oscillations \cite{NuData} 
motivate us to extend the SM so as to incorporate 
the neutrino masses and flavor mixings. 
One promising scenario is to introduce the right-handed neutrinos 
and their heavy Majorana masses, so that the tiny neutrino masses 
can be naturally explained by the seesaw mechanism \cite{seesawI}. 
By imposing the classical conformal invariance, the SM with the 
right-handed neutrinos has been recently investigated \cite{MaNi}. 
To keep the classical conformal invariance, a SM singlet scalar 
is introduced whose vacuum expectation value (VEV) generates 
masses of the right-handed neutrinos. 
Through numerical analysis, it has been shown that 
a set of parameters leads to phenomenologically viable results. 

In this letter, we investigate a $B-L$ gauged
extension of the SM with a classical conformal invariance.
In addition to the SM particles, 
the model has the right-handed neutrinos $\nu_R^i$
 and a SM singlet scalar $\Phi$ 
 whose VEV gives the Majorana mass terms for $\nu_R^i$
and also breaks the $B-L$ gauge symmetry.
Through an interaction with the scalar $\Phi$ field,
Higgs boson also acquires the VEV and the electroweak symmetry
is radiatively broken. 
Nevertheless, 
the dynamics of the SM Higgs sector is governed by the 
induced negative mass squared term, and if we impose
the triviality and the stability bounds on the Higgs potential, 
the Higgs boson mass is bounded in the same  range 
130 GeV$\lesssim m_h \lesssim$ 170 GeV as the
ordinary bounds in the SM \cite{Hmass}.
The phenomenological constraint from the neutrino mass
gives an upper bound for the $B-L$ breaking scale.

{\bf Model}.---
The model we will investigate is  the minimal $B-L$ extension of the SM 
\cite{B-L} with the classical conformal symmetry. 
The $B-L$  (baryon minus lepton) number is a unique anomaly free 
global symmetry that the SM accidentally possesses and 
can be easily gauged. 
Our model is based on the gauge group 
SU(3)$_c \times$SU(2)$_L\times$U(1)$_Y\times$U(1)$_{B-L}$ 
and the particle contents are listed in Table~1 \cite{B-L2}. 
Here, three generations of right-handed neutrinos ($\nu_R^i$)
are necessarily introduced to make the model free from all 
the gauge and gravitational anomalies. 
The SM singlet scalar ($\Phi$) works to break the U(1)$_{B-L}$ 
gauge symmetry by its VEV, and at the same time generates 
the right-handed neutrino masses. 
\begin{table}[t]
\begin{center}
\begin{tabular}{c|ccc|c}
            & SU(3)$_c$ & SU(2)$_L$ & U(1)$_Y$ & U(1)$_{B-L}$  \\
\hline
$ q_L^i $    & {\bf 3}   & {\bf 2}& $+1/6$ & $+1/3$  \\ 
$ u_R^i $    & {\bf 3} & {\bf 1}& $+2/3$ & $+1/3$  \\ 
$ d_R^i $    & {\bf 3} & {\bf 1}& $-1/3$ & $+1/3$  \\ 
\hline
$ \ell^i_L$    & {\bf 1} & {\bf 2}& $-1/2$ & $-1$  \\ 
$ \nu_R^i$   & {\bf 1} & {\bf 1}& $ 0$   & $-1$  \\ 
$ e_R^i  $   & {\bf 1} & {\bf 1}& $-1$   & $-1$  \\ 
\hline 
$ H$         & {\bf 1} & {\bf 2}& $-1/2$  &  $ 0$  \\ 
$ \Phi$      & {\bf 1} & {\bf 1}& $  0$  &  $+2$  \\ 
\end{tabular}
\end{center}
\caption{
Particle contents. 
In addition to the SM particle contents, 
the right-handed neutrino $\nu_R^i$ 
($i=1,2,3$ denotes the generation index) 
and a complex scalar $\Phi$ are introduced. 
}
\end{table}

The Lagrangian relevant for the seesaw mechanism is given as 
\bea 
 {\cal L} \supset -Y_D^{ij} \overline{\nu_R^i} H^\dagger \ell_L^j  
- \frac{1}{2} Y_N^i \Phi \overline{\nu_R^{i c}} \nu_R^i 
+{\rm h.c.},  
\label{Yukawa}
\eea
where the first term gives the Dirac neutrino mass term 
after the electroweak symmetry breaking, 
while the right-handed neutrino Majorana mass term 
is generated through the second term associated with 
the $B-L$ gauge symmetry breaking. 
Without loss of generality, we here work on the basis
where the second term is diagonalized and 
$Y_N^i$ is real and positive.

Under the hypothesis of the classical conformal invariance of the model, 
the classical scalar potential is described as 
\bea 
 V = \lambda_H (H^\dagger H)^2 + \lambda (\Phi^\dagger \Phi)^2 
   + \lambda^\prime (\Phi^\dagger \Phi) (H^\dagger H).  
\label{potential}
\eea
Note that when $\lambda^\prime$ is negligibly small, 
the SM Higgs sector and the $\Phi$ sector relevant 
for the $B-L$ symmetry breaking are approximately decoupled. 
If this is the case, we can separately analyze these two Higgs sectors. 
When the Yukawa coupling $Y_N$ is negligible compared to 
the U(1)$_{B-L}$ gauge coupling, the $\Phi$ sector is 
the same as the original Coleman-Weinberg model \cite{CW}, 
so that the radiative U(1)$_{B-L}$ symmetry breaking will be achieved. 
Once $\Phi$ develops its VEV, the tree-level mass term for the Higgs 
is effectively generated through the third term in Eq.~(\ref{potential}). 
Taking $\lambda^\prime$ negative, the induced mass squared for the Higgs 
doublet is negative and as a result, the electroweak symmetry breaking 
is driven in the same way as in the SM. 
In this letter, we show that in the $B-L$ extended SM with the classical 
conformal symmetry, such a simple symmetry breaking is realized 
in a wide range of the parameter space of the model 
which is consistent with both the theoretical and experimental 
requirements. 

{\bf $B-L$ symmetry breaking}.---
Assuming a negligible $\lambda^\prime$, let us first investigate 
the radiative $B-L$ symmetry breaking. 
We employ the renormalization group (RG) improved effective potential 
at the one-loop level of the form \cite{Sher} 
\bea 
 V (\phi) = \frac{1}{4} \lambda(t) G^4(t) \phi^4, 
\eea
where $\phi/\sqrt{2} =\Re[\Phi]$, $t=\log[\phi/M]$ 
with the renormalization point $M$, 
and 
\bea 
 G(t) = \exp \left[- \int_0^t d t^\prime \; 
 \gamma(t^\prime) \right] 
\eea 
with the anomalous dimension (in the Landau gauge) 
explicitly described as 
\bea
 \gamma = \frac{1}{32 \pi^2} 
 \left[ \sum_i (Y_N^i)^2 - a_2 g_{B-L}^2 \right].  
\eea
Here, $g_{B-L}$ is the $B-L$ gauge coupling, and $a_2=24$. 
Renormalization group equations for coupling parameters involved 
in our analysis are listed below: 
\bea 
2 \pi \frac{d \alpha_{B-L}}{d t} &=&  b \alpha_{B-L}^2, 
 \nonumber \\
2 \pi \frac{d \alpha_\lambda}{d t}&=& 
   a_1 \alpha_\lambda^2 + 8 \pi \alpha_\lambda \gamma 
 + a_3 \alpha_{B-L}^2 -\frac{1}{2} \sum_i (\alpha_N^i)^2, 
 \nonumber \\ 
\pi \frac{d \alpha_N^i}{d t} &=&  \alpha_N^i 
\left(
 \frac{1}{2} \alpha_N^i + \frac{1}{4} \sum_j \alpha_N^j - 9 \alpha_{B-L} 
\right), 
\label{RGEs}
\eea
where 
$\alpha_{B-L} = g_{B-L}^2/(4 \pi)$, $\alpha_\lambda = \lambda/(4 \pi)$, 
$\alpha_N^i = (Y_N^i)^2/(4 \pi)$, and the coefficients 
 in the beta functions are explicitly given as 
 $b=12$, $a_1=10$ and $a_3=48$.

Setting the renormalization point to be the VEV of $\phi$ 
at the potential minimum, $\phi=M$ or equivalently $t=0$, 
the stationary condition,  
\bea 
  \frac{d V}{d \phi} \Big|_{\phi=M}
= \frac{e^{-t}}{M} \frac{d V}{d t}\Big|_{t=0} = 0,  
\eea
leads to one condition among coupling constants 
at the potential minimum such that 
\bea 
\frac{d \alpha_\lambda}{d t} +4 \alpha_\lambda (1-\gamma) 
=\frac{1}{2 \pi}
\left(
10 \alpha_\lambda^2 + 48 \alpha_{B-L}^2 -\frac{1}{2} \sum_i
(\alpha_N^i)^2  \right) + 4 \alpha_\lambda =0  
\label{condition}
\eea
at $t=0$. 
For coupling values well within the perturbative regime, 
$\alpha_\lambda \sim \alpha_{B-L}^2 \sim (\alpha_N^i)^2  \ll 1$, 
we find a solution 
\bea 
 \alpha_\lambda(0) \simeq 
 -\frac{6}{\pi} 
  \left( 
 \alpha_{B-L}(0)^2 - \frac{1}{96} \sum_i (\alpha_N^i(0))^2 
  \right) . 
\label{solution}
\eea
In this approximation, it is straightforward to obtain 
the SM singlet Higgs boson mass of the form, 
\bea 
 m_\phi^2 = \frac{d^2 V}{d \phi^2}\Big|_{\phi=M}
 = \left( \frac{e^{-t}}{M} \frac{d}{dt}\right)^2 V\Big|_{t=0}
 \simeq - 16 \pi \alpha_\lambda(0) M^2
\label{phimass}  
\eea
under the conditions of Eq.~(\ref{solution}) 
and $\alpha_\lambda \sim \alpha_{B-L}^2 \sim (\alpha_N^i)^2  \ll 1$. 
Therefore, when we choose the coupling constant
to be $\alpha_\lambda(0) < 0$, 
 the effective potential has a minimum at $\phi=M$ and 
 the $B-L$ symmetry is radiatively 
broken\footnote{It may seem strange that the coupling constant
$\alpha_\lambda(0)$ must be taken negative 
for the extremum to be locally stable with a positive
mass squared. However the true 
effective coupling constant at the minimum 
should be defined as 
$\lambda_{eff}=V^{(4)}|_{t=0}/6$, which 
is related to $\lambda(0)$ as $\lambda(0)=-3 \lambda_{eff}/22$
and hence positive. Then the $\phi$ boson mass is given by the 
same formula as in 
(\ref{CWmass}).}.  
Note that in the limit $Y_N^i \to 0$, the system is the same 
as the one originally investigated by Coleman-Weinberg \cite{CW}, 
where the U(1) gauge interaction plays the crucial role 
to achieve the radiative symmetry breaking 
keeping the validity of perturbation. 
In this sense, gauging the U(1)$_{B-L}$ is necessary 
although it is not required for the purpose to implement 
the seesaw mechanism.

When $\alpha_N^i \ll \alpha_{B-L}$ and $\alpha_N^i$ is neglected, 
the analytic solutions of the RGEs in Eq.~(\ref{RGEs}) 
are known  \cite{CW}: 
\bea 
\alpha_{B-L}(t) &=& 
 \frac{\alpha_{B-L}(0)}{1-\frac{b}{2 \pi} \alpha_{B-L}(0) t}, 
\nonumber \\ 
\alpha_\lambda(t) &=& 
 \frac{a_2+ b}{2 a_1} \alpha_{B-L}(t) +
\frac{A}{a_1} \alpha_{B-L}(t)
 \tan\left[ \frac{A}{b} \ln \left( \frac{\alpha_{B-L}(t)}{\pi}\right)
 + C \right],
\label{exact}
\eea 
where 
 $A=\sqrt{a_1 a_3 -(a_2+b)^2/4}$, 
 and $C$ is the integration constant fixed to satisfy 
 Eq.~(\ref{condition}). 
Using these solutions, the RG improved effective potential 
is given by 
\bea 
 V = \pi 
\frac{\alpha_\lambda(t) M^4}{\left(1- \frac{b}{2 \pi} \alpha_{B-L}(0) t
\right)^{a_2/b}} e^{4 t}. 
\eea
We show the effective potential in Fig.~1 
for $\alpha_{B-L}=0.01$ as an example. 
Here $\alpha_\lambda(0)=- 1.91 \times 10^{-4}$ is fixed 
to satisfy Eq.~(\ref{condition}), and 
the effective potential has its minimum at $\phi/M=1$. 

The exact solution of $\alpha_\lambda(t)$ in Eq.~(\ref{exact}) 
becomes singular at some infrared region and thus 
the instability of the effective potential occurs \cite{MaNi2}. 
For $\alpha_{B-L}(0)=0.01$, for example, 
we find that this instability occurs at a very infrared point, 
$t \simeq -41$, and it is hard to recognize this point in Fig.~1. 
In a phenomenological point of view, this infrared point is 
far below the dynamical scale of the QCD and we cannot trust 
the effective potential from perturbative calculations. 
In this letter, we do not discuss this instability issue, but 
it is  interesting to note a connection of this instability
with the negativeness of $\alpha_\lambda(0)$. The latter
implies that the running coupling constant crosses zero
at some scale and the classical potential vanishes if
we choose the renormalization scale there. 
This is the condition for the perturbative 
validity of the Coleman-Weinberg scenario \cite{GilWein, Sher}.

\begin{figure}[ht]
\begin{center}
{\includegraphics*[width=.6\linewidth]{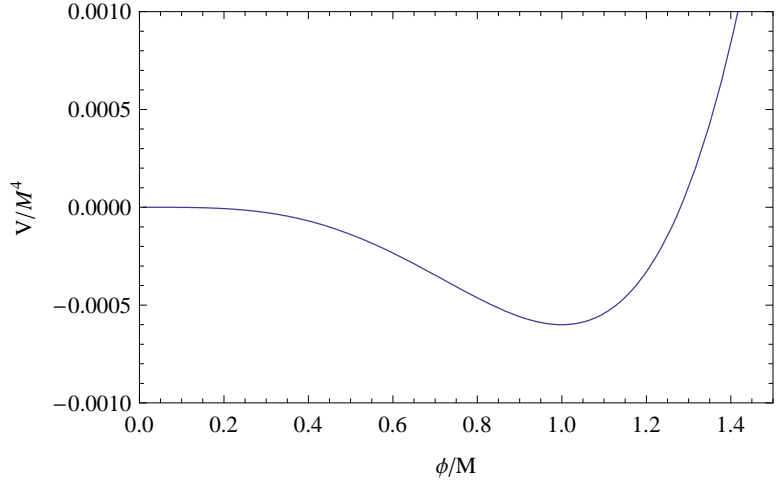}}
\caption{
The RG improved effective potential. 
Here, we have taken $\alpha_{B-L}(0)=0.01$, 
accordingly $\alpha_\lambda(0)=-1.91 \times 10^{-4}$ 
and $C = 1.27$ are fixed so that the effective potential 
has its minimum at $\phi/M=1$.
}
\end{center}
\end{figure}

An extra gauge boson associated with the U(1)$_{B-L}$ 
gauge symmetry, the so-called $Z'$ boson, acquires its mass 
through the $B-L$ symmetry breaking, $m_{Z'} = 2  
g_{B-L} M$. 
As is well-known, there is a relationship between the $Z'$ 
boson mass and the SM singlet Higgs boson mass 
in the Coleman-Weinberg model \cite{CW}. 
Neglecting $\alpha_N^i$ in Eqs.~(\ref{solution}) and (\ref{phimass}), 
 we find the mass relation 
\bea 
\left( \frac{m_\phi}{m_{Z'}} \right)^2 
 \simeq \frac{6}{\pi} \alpha_{B-L}.   
\label{phimass2}
\eea
The hierarchy between the two masses is a general consequence 
of the Coleman-Weinberg model where the symmetry breaking occurs 
under the balance between the tree-level quartic coupling and 
the terms generated by quantum corrections. 
The scalar boson $\phi$ can be much lighter than the $Z^\prime$
gauge boson and possibly comparable with the SM Higgs boson. 
Then, as we discuss later, the two scalars mix each other.

 Eqs.~(\ref{solution}) and (\ref{phimass}) suggest that 
as the Yukawa coupling $\alpha_N$ becomes larger, 
the SM singlet Higgs boson mass squared is reducing 
and eventually changes its sign. 
Therefore, there is an upper limit on the Yukawa coupling 
in order for the effective potential to have the minimum at $\phi=M$. 
This is in fact  the same reason as  
why the Coleman-Weinberg mechanism 
in the SM Higgs sector fails to break the electroweak symmetry 
when the top Yukawa coupling is large as observed. 
Analyzing the RG improved effective potential with only one Yukawa 
coupling $\alpha_N$, the SM singlet Higgs boson mass as a function 
of the Yukawa coupling is depicted in Fig.~2. 
The minimum at $M$ in the effective potential changes into 
the maximum for $\alpha_N(0) > 9.8 \alpha_{B-L}(0)$. 

\begin{figure}[ht]
\begin{center}
{\includegraphics*[width=.6\linewidth]{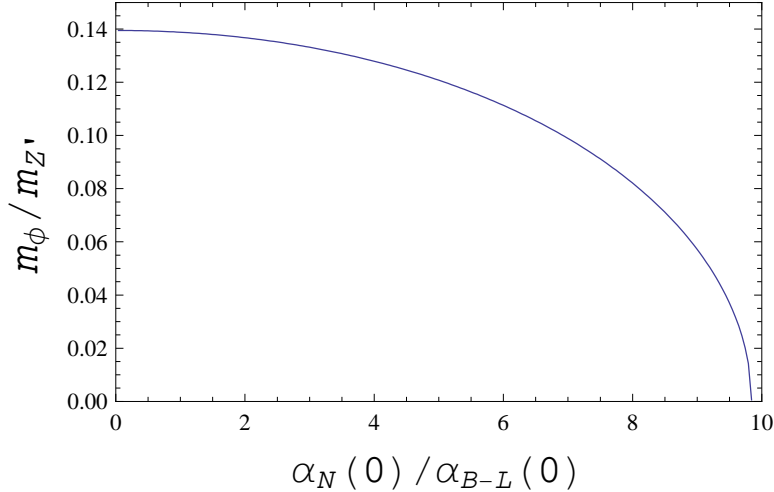}}
\caption{
The SM singlet Higgs boson mass as a function of the Yukawa coupling. 
Here we have taken $\alpha_{B-L}(0)=0.01$ and accordingly,
fixed $\alpha_\lambda(0)$ to satisfy the stationary condition 
in Eq.~(\ref{condition}). 
For $\alpha_N(0) \simeq 9.8 \alpha_{B-L}(0)$, the potential minimum 
at $\phi=M$ changes into the maximum. 
}
\end{center}
\end{figure}
{\bf Electroweak symmetry breaking}.---
Now let us consider the SM Higgs sector. 
In our model, the electroweak symmetry breaking is achieved 
in a very simple way. 
Once the $B-L$ symmetry is broken, the SM Higgs doublet mass 
is generated through the mixing term between $H$ and $\Phi$ 
in the scalar potential (see Eqs.~(\ref{potential}) and 
(\ref{classicalpotential})), 
\bea 
  \mu^2 = \frac{\lambda^\prime}{2} M^2.  
\label{Hmass}
\eea
Choosing $\lambda^\prime < 0$, the electroweak symmetry is 
broken in the same way as in the SM. 
However, the crucial difference from the SM is that 
in our model, the electroweak symmetry breaking originates form 
the radiative breaking of the U(1)$_{B-L}$ gauge symmetry. 
At the tree level the Higgs boson mass 
is given by $m_h^2=2 |\mu^2|=|\lambda^\prime|M^2=
2 \lambda_H v^2$
where $\langle h \rangle = v =246$ GeV. 
Note that the mass is independent of $\lambda^\prime$
when it is written in terms of the VEV of $h$.
If $\lambda^\prime$ is sufficiently small, 
the mass formula for the Higgs boson is given by Eq.(\ref{SMmass}).
Then, by imposing the triviality (up to the Planck scale)
and the vacuum stability bounds, 
the Higgs boson mass is given in a range 
130 GeV$\lesssim m_h \lesssim$ 170 GeV  as in the 
SM \cite{Hmass}. 
When we include the effect of 
the mixing with the $B-L$ sector, 
the Higgs boson mass has a correction (see Eq.(\ref{CWSMmass}))  
\begin{equation}
\Delta m_h^2 = -\frac{64}{3} B_\phi v^2,
\end{equation}
where $B_\phi$ is the 1-loop coefficient of the $\phi$ field
contribution to the Higgs potential;
\begin{equation}
B_\phi = \frac{\lambda^{\prime 2}}{128 \pi^2}.
\end{equation}
The correction is negligible if we take $|\lambda^\prime|$ 
as a phenomenologically favorable value (discussed later),
e.g. $\lambda' \sim 0.002$ for $M \sim 3$ TeV.
Since $|\lambda^\prime|=2 \lambda_H (v/M)^2$,
it becomes smaller for larger $M$.
The effect of the right-handed neutrinos to the coefficient $B$ 
is also roughly given by $Y_D^4/(16 \pi^2)$, which can be sizable 
when the right-handed neutrino mass scale is higher than 
$10^{13}$ GeV \cite{SSeffect1, SSeffect2}. 
In our case, the effect is negligible 
since the right-handed neutrino mass scale should be 
small $\ll 10^{13}$ GeV as will be discussed later. 
Another effect of the right-handed neutrinos to the coefficient 
$\lambda^\prime$ is also discussed later.

{\bf Validity of the approximation}.---
We have discussed the $B-L$ symmetry breaking sector and 
the SM Higgs sector separately and in this case, 
both the symmetry breakings are easily realized. 
This treatment 
should be justified only in a limited parameter 
space of the model, but we will see that
the validity of this approximation can be held
easily once we impose a phenomenologically 
viable condition for the $B-L$ symmetry breaking
scale $M>$ a few TeV.

First, we determine the upper bound of the $B-L$ gauge coupling.
We require all the coupling constants 
in our model are in the perturbative regime below 
the Planck scale. 
Because of its large beta function coefficient, 
the $B-L$ gauge coupling is severely constrained
by this condition. 
We can find that 
$\alpha_{B-L}(0) \lesssim 0.015$ is required 
if we take its renormalization scale at $M \gtrsim 3$ TeV. 
(The experimental lower bound for the $Z^\prime$ gauge boson
requires $M$  to be larger than this value as discussed later.)

Now we will investigate the upper bound
for the coupling constant $\lambda'$ 
in order to satisfy the separability condition for 
the $B-L$  and the electroweak sectors.
Using $\phi$ as the background field for $H$, 
the negative Higgs doublet mass 
squared is induced and as a result, the electroweak symmetry 
is broken. 
Once the Higgs doublet develops a VEV, $\phi$ 
is mixed with the Higgs field $h$
through the coupling $\lambda^\prime$.
Around the minimum $(h=v, \phi=M)$, 
the condition that the mixing term does not 
drastically deform the potential of $\phi$ with a curvature 
$m_\phi^2$  is given by
\bea 
 |\lambda^\prime| M v \ll m_\phi^2. 
\eea
Neglecting $Y_N^i$, again, in Eq.~(\ref{phimass}), for simplicity, 
we find 
\bea 
 |\lambda^\prime| \ll 96 \alpha_{B-L}^2 
 \left(\frac{M}{v}\right).  
\eea 
For example, if we take $\alpha_{B-L}=0.01$ and $M=3$ TeV,
the condition becomes 
$|\lambda^\prime| \ll 0.12 $. 
As $M$ becomes larger, the constraint on $\lambda^\prime$ becomes looser.

Another condition we need to check is the 
smallness of the $\lambda^\prime$ contribution to the beta 
function of $\alpha_{\lambda}$, which we have neglected.
Non-zero $\lambda^\prime$ contributes to the beta function of 
$\lambda$ and adds a term $\alpha_{\lambda^\prime}^2$ 
in the right hand side of Eq.~(\ref{RGEs}), 
where $\alpha_{\lambda^\prime}= \lambda^\prime/(4 \pi)$. 
In order not to change our analysis of the scalar potential
of $\phi$, let us require 
\bea 
 \alpha_{\lambda^\prime}^2 \ll 48 \alpha_{B-L}^2 .
\eea
We find $|\lambda^\prime| \ll 0.87 $ for $\alpha_{B-L}=0.01$. 
It is looser than the first condition.

The final, but the most stringent  condition for $\lambda'$ comes from the
Higgs boson mass bounds.
The SM Higgs potential with the induced Higgs doublet mass term 
in Eq.~(\ref{Hmass}) leads to the relation  
\bea 
  |\lambda^\prime| =\left( \frac{m_h}{M} \right)^2,  
\eea
at the electroweak symmetry breaking vacuum, 
where $m_h$ is the physical Higgs boson mass.
Since the electroweak symmetry breaking is triggered by
the  negative mass term, the dynamics is controlled 
by the classical action in Eq. (\ref{classicalpotential}). 
Therefore,
the triviality and the vacuum stability bounds in the SM \cite{Hmass} 
require the Higgs boson mass in the range 
130 GeV$\lesssim m_h \lesssim$ 170 GeV. 
Since for the phenomenologically favorable value of 
$M \gtrsim$ 3 TeV, 
the contribution of $\lambda^\prime$ 
to the RGE evolution of the Higgs quartic coupling $\lambda_H$ 
is negligible, we can apply this Higgs boson mass bound in the SM 
to constrain the range of $\lambda^\prime$ such as 
$(130 {\rm GeV}/M)^2 \lesssim |\lambda^\prime| 
\lesssim (170 {\rm GeV}/M)^2$. 
This bound on $\lambda^\prime$ for $M \gtrsim 3$  TeV 
becomes $0.0019 \lesssim |\lambda^\prime| \lesssim 0.0032$
and 
is consistent with those evaluated above for $\alpha_{B-L}=0.01$. 
The larger value of $M$ reduces $\lambda^\prime$ and the 
two sectors can be further treated separately.

\begin{figure}[t]\begin{center}
\includegraphics[scale=1.2]{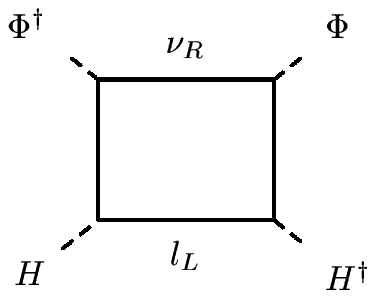}
\caption{
One-loop diagram inducing the mixing term 
 $(\Phi^\dagger \Phi)(H^\dagger H )$ 
 through the right-handed neutrinos. 
}
\end{center}
\end{figure}
\begin{figure}[t]\begin{center}
\includegraphics[scale=1.2]{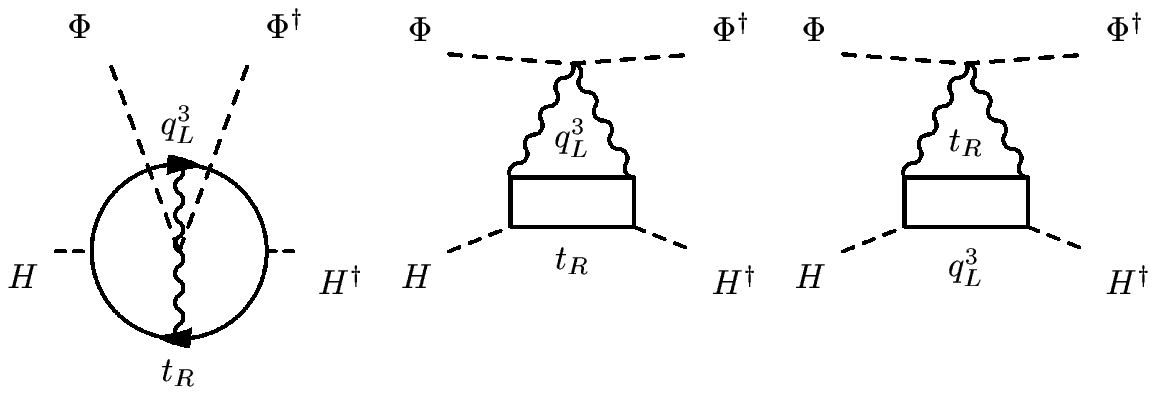}
\caption{
Two-loop diagrams inducing the mixing term $(\Phi^\dagger \Phi)(H^\dagger H )$ 
 through the top-quarks and the $B-L$ gauge bosons. 
 The wavy lines represent the 
 propagators of the $B-L$ gauge bosons.
}
\end{center}
\end{figure}

{\bf Scale of the $B-L$ breaking}.---
The energy scale of the $B-L$ symmetry breaking is  constrained
from both of the above and the below
by the stability of the electroweak scale and
the experimental bound for the $Z^\prime$ gauge boson mass.

We have imposed the classical conformal invariance 
and the absence of the quadratic divergences to solve
the gauge hierarchy problem, but 
if the $B-L$ symmetry breaking scale is large, 
we may need a fine tuning of the electroweak scale
again to cancel the radiative corrections by some heavy 
states associated  with the $B-L$ symmetry breaking\footnote{
A similar discussion is given in \cite{Casas}.
}.
At the one-loop level (see Fig.~3), 
the mixing term $(\Phi^\dagger \Phi)(H^\dagger H)$ can be also 
induced through the right-handed neutrinos 
with the Yukawa interactions in Eq.~(\ref{Yukawa}). 
Substituting $\phi=M$, we obtain the correction to 
the effective Higgs boson mass squared such as 
\bea 
 \Delta m_h^2 \sim  \frac {Y_D^2 Y_N^2}{16 \pi^2}M^2 
 \sim \frac {m_\nu M_N^3}{16 \pi^2 v^2} ,
\eea
where we have used the seesaw formula, 
$m_\nu \sim Y_D^2 v^2/M_N$ with $M_N = Y_N M$. 
For the stability of the electroweak vacuum, 
$\Delta m_h^2$ should be smaller than the electroweak scale\footnote{
Note that the condition is equivalent
to an issue of the {\it naturalness} of the $\lambda^\prime$ coupling.
The dynamics of the SM Higgs sector is still governed
by the classical potential with a negative mass squared, and
hence the Higgs boson mass bounds are not modified by this type of the
corrections.
}. 
Thus, we can obtain the upper bound of $M_N$ once $m_\nu$ is fixed. 
For example, when the neutrino mass is around $m_\nu \sim 0.1$ eV,
there is an upper bound for the  Majorana mass
$M_N \lesssim 10^7$ GeV and hence $ M \lesssim 10^7/Y_N$ GeV. 
As we discussed  before, the Yukawa coupling $Y_N$ cannot
be larger than the $B-L$ gauge coupling $g_{B-L}$ 
in order to realize the radiative breaking in the $B-L$ sector. 

We also consider higher order corrections to the mixing  term, 
which are dominated by two-loop diagrams involving top-quarks 
and the $B-L$ gauge boson (see Fig.~4). 
Again, substituting $\phi=M$, we obtain the correction such as 
\bea 
 \Delta m_h^2 \sim Y_t^2 \left(\frac{\alpha_{B-L}}{4 \pi} \right)^2 M^2,     
\eea
where $Y_t \sim 1$ is the top quark Yukawa coupling. 
This correction leads to the upper bound 
 $M \lesssim 10^5 \; {\rm GeV}\times (0.01/\alpha_{B-L})$, 
 which is severer than the correction given by the right-handed neutrino. 
If there exist other mass scales, additional radiative contributions to 
 the Higgs boson mass can be generated. 
Thus, in order to ensure the stability of the electroweak scale, 
 we implicitly assume that there exists no other new physics 
 between the $B-L$ breaking scale and the Planck scale.

On the other hand, there is an experimental lower bound for the scale $M$.  
The $Z^\prime$ boson with mass $m_{Z^\prime} = 2 
g_{B-L} M$ 
couples to all the SM fermions as well as the right-handed neutrinos 
through the $B-L$ charges. 
The experimental search for the $Z^\prime$ boson at LEP II 
gives the limit on the $B-L$ symmetry breaking scale 
$ M  \gtrsim 3$ TeV \cite{Zbound}. 
This bound is consistent with the bound from the direct search 
of the $Z^\prime$ boson at Tevatron \cite{CDF}. 

{\bf Conclusions and discussions}.---
The hypothesis of the classical conformal invariance 
may solve the gauge hierarchy problem in the Standard Model 
\cite{Bardeen}. 
Once this hypothesis is imposed on gauge theories, 
the gauge symmetry can be radiatively broken in 
the Coleman-Weinberg potential. 
Unfortunately, this scenario cannot work in the Standard Model 
since the large top Yukawa coupling destabilizes the effective 
Higgs potential and hence, some extensions of the model 
are necessary. 
Motivated by the recent observations of neutrino masses 
and flavor mixings, we consider the minimal $B-L$ extension 
of the Standard Model, in which the right-handed neutrinos 
of each generation are necessarily introduced to cancel 
the gauge anomalies and tiny neutrino masses can be naturally 
explained by the seesaw mechanism. 
Under the hypothesis of the classical conformal invariance, 
we have investigated the radiative gauge symmetry breaking 
in this model. 
Taking the mixing parameter between the SM Higgs and the SM singlet 
scalars to be small, we have analyzed the effective potential 
of the Higgs and the singlet sectors separately and proposed a very simple 
realization of both the $B-L$ and the 
Standard Model gauge symmetry breakings. 
We have found that this scenario can work for a wide range of 
parameter space keeping the theoretical and phenomenological requirements.

Finally, we comment on phenomenological aspects of our model. 
If the $B-L$ symmetry breaking scale is around its experimental 
lower bound $M \simeq 3$ TeV,  
all new particles in the $B-L$ extended SM, 
the $Z^\prime$ boson, the right-handed neutrinos and 
the SM singlet Higgs boson, can be as light as a few TeV. 
There has been a number of works of 
Large Hadron Collider (LHC) phenomenologies 
on the production of such new particles \cite{LHC}. 
In particular, the $Z^\prime$ boson can be produced 
through Drell-Yang processes and once produced, 
its decay into di-leptons with a large branching ratio 
would provide us clean signatures.

After the electroweak symmetry breaking, 
the SM singlet Higgs boson mixes with the SM Higgs boson. 
In our model the SM singlet Higgs boson is relatively light 
to the $Z^\prime$ boson (see Eq.~(\ref{phimass2})) and 
its mass can be as low as a few hundred GeV. 
When the singlet Higgs boson is light, the mixing between 
two Higgs bosons can be sizable and affect 
the SM Higgs boson phenomenology.

The leptogenesis \cite{F-Y} through the lepton number 
and CP violating decays of the right-handed Majorana neutrino 
is a very simple mechanism for baryogenesis. 
In normal thermal leptogenesis scenario, there is 
a lower mass bound on the lightest right-handed neutrino, 
$M_N \gtrsim 10^9$ GeV \cite{LowerBound}, 
in order to achieve the realistic baryon asymmetry 
of the present universe. 
In our model, as discussed above, the theoretical requirement 
constrains the right-handed neutrino mass 
to be smaller than this bound, for example, 
$M_N \lesssim Y_N \times 10^5$ GeV for $\alpha_{B-L} \sim 0.01$. 
However, the leptogenesis is still possible through 
the resonant leptogenesis \cite{resonantLG} 
when right-handed neutrino masses are well-degenerated. 
One interesting feature in our model is that at high temperature, 
the right-handed neutrinos can be in thermal equilibrium 
with the SM particles through the $B-L$ gauge interaction, 
so that a large efficiency factor can be easily obtained.

The existence of (cold) dark matter is strongly supported 
by various observations of the present universe 
and suggests the need of extending the SM 
since the SM includes no suitable candidate for dark matter. 
A simply extended model has been proposed \cite{DMmodel1, DMmodel2}, 
where only a SM singlet scalar with an odd parity is introduced. 
This scalar can be a suitable candidate for dark matter 
through the coupling with the SM Higgs boson. 
It is easy to extend our model in the same way, i.e., 
we introduce a parity odd scalar which is singlet under 
both the SM and $B-L$ gauge groups.  
According to our hypothesis of the classical conformal invariance, 
this singlet scalar has no mass term. 
There are only two terms for the interaction 
among the dark matter and the other particles 
at the renormalizable level, 
%
\bea 
 {\cal L} \supset 
 -\lambda_1 (H^\dagger H) \chi^2 - \lambda_2 (\Phi^\dagger \Phi) \chi^2,  
\eea 
where $\chi$ is the dark matter candidate. 
The mass for the dark matter is generated 
by the $B-L$ and electroweak symmetry breakings, 
$m_\chi^2 = \lambda_1 v^2 + \lambda_2 M^2$, 
together with the interaction terms 
among the dark matter and the Higgs bosons, 
\bea
 {\cal L}_{int} \supset  
  -\frac{\lambda_1}{2} (2 v h + h^2) \chi^2 
  -\frac{\lambda_2}{2} (2 M \phi + \phi^2) \chi^2,  
\eea
which plays the essential role in the dark matter annihilation 
processes in the early universe. 
When we neglect $\lambda_2$, this model is essentially 
the same as the one discussed in \cite{DMmodel2}. 
Thanks to the absence of the dark matter  mass term, 
dark matter physics is controlled by only 
two unknown parameters, $m_\chi$ and $m_h$, 
which are well constrained by the observed 
dark matter  relic density \cite{DMmodel2}. 
The couplings $\lambda_1$ and $\lambda_2 $ are involved in 
the RGEs of quartic Higgs couplings and potentially affect 
the SM Higgs boson mass bound from triviality (up to the Planck scale) 
and vacuum stability arguments. 
We leave this study for future works.

\section*{Acknowledgments}
We would like to thank Hajime Aoki,  Eung Jin Chun and Kazuo Fujikawa
for useful discussions and comments. 
This work of N.O. is supported in part by the Grant-in-Aid 
for Scientific Research from the Ministry of Education, 
Science and Culture of Japan, No.~18740170.




\begin{thebibliography}{99}

\bibitem{Bardeen}
W.~A.~Bardeen,
  FERMILAB-CONF-95-391-T 


\bibitem{CW}
S.~R.~Coleman and E.~J.~Weinberg,
  Phys.\ Rev.\  D {\bf 7}, 1888 (1973).


\bibitem{LEP2}
R.~Barate {\it et al.}  [LEP Working Group for Higgs boson searches],
 Phys.\ Lett.\  B {\bf 565}, 61 (2003).


\bibitem{Tevatron}
[Tevatron Electroweak Working Group and CDF Collaboration and D0 Collab],
  arXiv:0808.1089 [hep-ex].


\bibitem{fujikawa}
 K.~Fujikawa,
  Prog.\ Theor.\ Phys.\  {\bf 61}, 1186 (1979).


\bibitem{Hempfling}
R.~Hempfling,
  Phys.\ Lett.\  B {\bf 379}, 153 (1996)
  [arXiv:hep-ph/9604278].

\bibitem{Espinosa}
J.~R.~Espinosa and M.~Quiros,
  Phys.\ Rev.\  D {\bf 76}, 076004 (2007)
  [arXiv:hep-ph/0701145].

\bibitem{Chang}
W. F. Chang, J. N. Ng and J. M. S. Wu,
  Phys.\ Rev.\ D {\bf 75}, 115016 (2007)
  [arXiv:hep-ph/0701254].

\bibitem{Foot}
R.~Foot, A.~Kobakhidze and R.~R.~Volkas,
  Phys.\ Lett.\  B {\bf 655}, 156 (2007)
  [arXiv:0704.1165 [hep-ph]]; 
%
R.~Foot, A.~Kobakhidze, K.~L.~McDonald and R.~R.~Volkas,
  Phys.\ Rev.\  D {\bf 76}, 075014 (2007)
  [arXiv:0706.1829 [hep-ph]]; 
%
  Phys.\ Rev.\  D {\bf 77}, 035006 (2008)
  [arXiv:0709.2750 [hep-ph]].


\bibitem{MaNi}
K.~A.~Meissner and H.~Nicolai,
  Phys.\ Lett.\  B {\bf 648}, 312 (2007)
  [arXiv:hep-th/0612165]; 
%
  Phys.\ Lett.\  B {\bf 660}, 260 (2008)
  [arXiv:0710.2840 [hep-th]]; 
%
  Eur.\ Phys.\ J.\  C {\bf 57}, 493 (2008)
  [arXiv:0803.2814 [hep-th]].


\bibitem{NuData}
B. T. Cleveland {\it et.al}, Astrophys.J. {\bf 496} 505 (1998);
%
Super-Kamiokande Collaboration, Phys. Lett. {\bf B539} 179 (2002);
Super-Kamiokande Collaboration, Phys. Rev. {\bf D71} 112005 (2005);
%
M. Maltoni, T. Schwetz, M.A. Tortola, J.W.F. Valle
 New J.Phys. {\bf 6} 122 (2004);
A. Bandyopadhyay {\it et al},
 Phys. Lett. {\bf B608} 115 (2005);
G. L. Fogli {\it et al},
 Prog. Part. Nucl. Phys. {\bf 57} 742 (2006);
For a recent review, see, for example,
H.~Nunokawa, S.~J.~Parke and J.~W.~F.~Valle,
 Prog.\ Part.\ Nucl.\ Phys.\  {\bf 60}, 338 (2008).


\bibitem{seesawI}
P.~Minkowski, Phys. Lett. B {\bf 67}, 421 (1977);
T.~Yanagida, in \emph{Proceedings of the Workshop on the Unified
  Theory and the Baryon Number in the Universe} (O.~Sawada and
  A.~Sugamoto, eds.), KEK, Tsukuba, Japan, 1979, p.~95;
M.~Gell-Mann, P.~Ramond, and R.~Slansky, \emph{Supergravity} (P.~van
  Nieuwenhuizen et al. eds.), North Holland, Amsterdam, 1979, p.~315;
S.~L. Glashow, \emph{The future of elementary particle physics}, in
  \emph{Proceedings of the 1979 Carg{\`e}se Summer Institute
 on Quarks and Leptons} (M.~Levy et al. eds.),
 Plenum Press, New York, 1980, p.~687;
R.~N. Mohapatra and G.~Senjanovic,
 Phys. Rev. Lett. {\bf 44}, 912 (1980).


\bibitem{Hmass}
U.~Ellwanger, Phys.\ Lett.\ {\bf B 303} (1993) 271;
U.~Ellwanger, M.~Rausch de Traubenberg and C.~A.~Savoy,
 Phys.\ Lett.\ {\bf B 315} (1993) 331;
T.~Elliott, S.~F.~King and P.~L.~White,
 Phys.\ Lett.\ {\bf B351} (1995) 213;
S.~F.~King and P.~L.~White,
 Phys.\ Rev.\ {\bf D52} (1995) 4183 ;
  G.~K.~Yeghian,
  arXiv:hep-ph/9904488;
   U.~Ellwanger and C.~Hugonie,
  Mod.\ Phys.\ Lett.\  A {\bf 22}, 1581 (2007).


\bibitem{B-L}
R.~N.~Mohapatra and R.~E.~Marshak,
  Phys.\ Rev.\ Lett.\  {\bf 44}, 1316 (1980);
R.~E.~Marshak and R.~N.~Mohapatra,
  Phys.\ Lett.\  B {\bf 91}, 222 (1980);
C.~Wetterich,
  Nucl.\ Phys.\  B {\bf 187}, 343 (1981);
A.~Masiero, J.~F.~Nieves and T.~Yanagida,
  Phys.\ Lett.\  B {\bf 116}, 11 (1982);
R.~N.~Mohapatra and G.~Senjanovic,
  Phys.\ Rev.\  D {\bf 27}, 254 (1983);
W.~Buchmuller, C.~Greub and P.~Minkowski,
  Phys.\ Lett.\  B {\bf 267}, 395 (1991). 


\bibitem{B-L2}
S.~Khalil,
  J.\ Phys.\ G {\bf 35}, 055001 (2008)
  [arXiv:hep-ph/0611205].


\bibitem{Sher} 
For a review, see 
M.~Sher,
  Phys.\ Rept.\  {\bf 179}, 273 (1989), 
and references therein. 


\bibitem{MaNi2}
K.~A.~Meissner and H.~Nicolai,
  arXiv:0809.1338 [hep-th].


\bibitem{GilWein}
E.~Gildener and S.~Weinberg,
  Phys.\ Rev.\  D {\bf 13}, 3333 (1976).


\bibitem{SSeffect1}
J.~A.~Casas, V.~Di Clemente, A.~Ibarra and M.~Quiros,
  Phys.\ Rev.\  D {\bf 62}, 053005 (2000)
  [arXiv:hep-ph/9904295].


\bibitem{SSeffect2}
I.~Gogoladze, N.~Okada and Q.~Shafi,
  Phys.\ Lett.\  B {\bf 668}, 121 (2008)
  [arXiv:0805.2129 [hep-ph]].


\bibitem{Casas}
  J.~A.~Casas, J.~R.~Espinosa and I.~Hidalgo,
  JHEP {\bf 0411}, 057 (2004)
  [arXiv:hep-ph/0410298].


\bibitem{Zbound}
M.~S.~Carena, A.~Daleo, B.~A.~Dobrescu and T.~M.~P.~Tait,
  Phys.\ Rev.\  D {\bf 70}, 093009 (2004)
  [arXiv:hep-ph/0408098].


\bibitem{CDF}
A.~Abulencia {\it et al.}  [CDF Collaboration],
  Phys.\ Rev.\ Lett.\  {\bf 96}, 211801 (2006)
  [arXiv:hep-ex/0602045].


\bibitem{LHC}
W.~Emam and S.~Khalil,
  Eur.\ Phys.\ J.\  C {\bf 522}, 625 (2007)
  [arXiv:0704.1395 [hep-ph]]; 
%
M.~Abbas and S.~Khalil,
  JHEP {\bf 0804}, 056 (2008)
  [arXiv:0707.0841 [hep-ph]].
%
M.~Abbas, W.~Emam, S.~Khalil and M.~Shalaby,
  Int.\ J.\ Mod.\ Phys.\  A {\bf 22}, 5889 (2008); 
%
K.~Huitu, S.~Khalil, H.~Okada and S.~K.~Rai,
  Phys.\ Rev.\ Lett.\  {\bf 101}, 181802 (2008)
  [arXiv:0803.2799 [hep-ph]]; 
%
L.~Basso, A.~Belyaev, S.~Moretti and C.~H.~Shepherd-Themistocleous,
  arXiv:0812.4313 [hep-ph].


\bibitem{F-Y}
M.~Fukugita and T.~Yanagida,
 Phys.\ Lett.\ B {\bf 174}, 45 (1986).


\bibitem{LowerBound}
W.~Buchmuller, P.~Di Bari and M.~Plumacher,
  Nucl.\ Phys.\  B {\bf 643}, 367 (2002)
  [Erratum-ibid.\  B {\bf 793}, 362 (2008)]
  [arXiv:hep-ph/0205349].


\bibitem{resonantLG} 
M.~Flanz, E.~A.~Paschos, U.~Sarkar and J.~Weiss,
  Phys.\ Lett.\  B {\bf 389}, 693 (1996)
  [arXiv:hep-ph/9607310]; 
%
A.~Pilaftsis,
  Phys.\ Rev.\  D {\bf 56}, 5431 (1997)
  [arXiv:hep-ph/9707235]; 
%
A.~Pilaftsis and T.~E.~J.~Underwood,
  Nucl.\ Phys.\  B {\bf 692}, 303 (2004)
  [arXiv:hep-ph/0309342].


\bibitem{DMmodel1}
J.~McDonald,
  Phys.\ Rev.\  D {\bf 50}, 3637 (1994)
 [arXiv:hep-ph/0702143]; 
%
C.~P.~Burgess, M.~Pospelov and T.~ter Veldhuis,
  Nucl.\ Phys.\  B {\bf 619}, 709 (2001)
  [arXiv:hep-ph/0011335]; 
%
M.~C.~Bento, O.~Bertolami and R.~Rosenfeld,
  Phys.\ Lett.\  B {\bf 518}, 276 (2001)
  [arXiv:hep-ph/0103340];
%
H.~Davoudiasl, R.~Kitano, T.~Li and H.~Murayama,
  Phys.\ Lett.\  B {\bf 609}, 117 (2005)
  [arXiv:hep-ph/0405097].


\bibitem{DMmodel2}
T.~Kikuchi and N.~Okada,
  Phys.\ Lett.\  B {\bf 665}, 186 (2008)
  [arXiv:0711.1506 [hep-ph]].



\end{thebibliography}
\end{document}